\documentstyle[12pt,psfig]{article}
\def\ref{\par\noindent\hangindent=6mm\hangafter=1}
\rm
\begin{document}
\baselineskip=7mm

\begin{center}{\large\bf Relation between millimeter wavelengths emission and
high-energy emission for active galactic nuclei }
\end{center}

\begin{center}
{\bf L. H. Huang, D. R. Jiang, Xinwu Cao\\
Shanghai Observatory, Chinese Academy of Sciences, 80 Nandan Road,\\
Shanghai, 200030, P.R. China}
\end{center}

\vskip 10mm
\centerline{to appear in {\it Chinese Physics Letters}}
\vskip 10mm

\centerline{\bf Abstract}

 After comparing the flux densities of a sample of active galactic nuclei
 detected by energetic gamma-ray experiment telescope at 90 
and 230 GHz with the $\gamma$-ray emissions detected by Compton
Gamma Ray Observatory and x-ray
emission, a strong correlation between the emissions
at the millimeter wavelength and the $\gamma$-ray emission is found.
The average flux density of x-ray is almost proportional to the average flux
density at the millimeter wavelength for quasars detected by
energetic gamma-ray experiment telescope, which
strongly supports the previous idea that the x-ray emissions of this kind
sources are
mainly produced by Synchrotron Self-Compton process.


\newpage
The identification of more than fifty EGRET (energeric gamma ray experiment
telescope)sources with radio-loud active
galactic nuclei (AGNs)
has drawn the attention
of astronomical community to these very interesting extragalactic objects $^1$.
Nearly all of these AGNs are characterized by one or more of the following
emission features: flat spectrum, core-dominated, superluminal radio
emission, rapid optical and $\gamma $-ray variability, and high optical
polarization. These properties are believed to be heavily influenced by
the relativistic beaming in a narrow jet with near light speed velocities
directed at a small angle to the line of sight.
Many models have been proposed to explain the
origin of high energy emission, including the two basic models: synchrotron
self-Compton (SSC) model and external Compton model, where seed
photons might come from accretion disc, broad line region or the dusty torus.
Comparisons with 22  and 37 GHz radio monitoring
data indicated that many of EGRET sources were in an enhanced radio state at
the time of $\gamma $-ray detection $^2$
 In this paper, we will concentrate on a sample of
EGRET sources detected so far (Table 1).
The flux densities at 90 and 230 GHz come mainly from Tornikoski
et al $^3$, Steppe et al $^4$ and Reuter et al $^5$, where a large number
of observations at millimeter wavelengths are available. The $\gamma $-ray data
come mainly from Mukherjee et al $^1$, where all EGRET observations from
phase 1 to phase 4 are available. So, we do not list them in Table 1. The
average flux densities over all observations at 90 and 230 GHz are
listed in Table 1.
The x-ray data and their references are also listed in Table 1. The spectral
index at millimeter wavelengths is referred to Bloom et al $^6$ and
Gear et al $^7$.
We K-correct the observed fluxes according to:
\begin{equation}
F(\nu)=F_{obs}(\nu)(1+z)^{\alpha-1}
\end{equation}
where $\alpha$ is the spectral index at the frequency
$\nu$[$F(\nu)\propto{\nu}^{-\alpha}$]. We use $\alpha$=1 in the x-ray band and
$\alpha$=0.5 in the millimeter waveband when the spectral index is not available.
In Fig. 1(a), we plot the average 90 GHz flux against the high state of $\gamma$-ray
emission for 29 EGRET AGNs for which there were 90 GHz data within 0.2 yr
of the $\gamma$-ray observation. Figure 2(b) corresponds to the case of 230 GHz.
Table 2 lists the correlation results  using linear regression analysis.
A strong correlation between the $\gamma$-ray emission and the emission at
millimeter wavelength
is revealed. This is consistent with that the high levels of $\gamma$-ray
emission are
associated with the millimeter flares, which has been found in several famous sources
such as 3C279 $^1$$^5$ , 0528+134 $^1$$^6$, 1730-130 $^1$$^7$ and in most of the
sources of our sample.
Figure 1(a) shows that two sources (1156+295, 3C273)deviate from the
correlation a lot.
The 1156+295 was observed to reach a historical maximun in early 1993 but there
were few millimeter observations during the time, the highest flux at 90 GHz
might not be recorded. The strongest radio source 3C273 might have its
$\gamma$-ray
emission concentrated on MeV, which results in a low $\gamma$-ray flux
detected by EGRET.
Figures 1(c) and 1(d) show the average 90 GHz flux density versus the x-ray flux density
and the average 230
GHz flux density versus the x-ray flux density, respectively. The flux densities
are
averaged over all observations since it is difficult to find the
quasi-simultaneous observations.
For EGRET QSOs, a strong correlation is
derived between the average x-ray flux and millimeter wavelength flux.
The
relation between the fluxes at 90 GHz , 230 GHz and the flux of x-ray
is  also listed in the Table 2.
For 3C273,
a correlated variability between the x-ray emission and the emission
at millimeter wavelength was obtained by McHardy et al $^1$$^8$
, and similar result was recently reported by Wehrle et al for
3C279. All these observations are consistent with our 
statistical result.
When BL Lacs are included, the correlations
become worse.
This implies that the x-ray emission could not be attributed to SSC for BL Lac
objects. This can also be shown by the
difference between the
spectral index of BL Lacs and that of flat radio spectrum quasars (FRSQs): $<\alpha _x>=1.43\pm 0.21$ for
BL Lacs and $<\alpha _x>=0.67\pm 0.13$ for EGRET FRSQs $^9$.
The average spectral index at millimeter-submillimeter wavelength is
0.75 for all FRSQs and about 0.6 for EGRET FRSQs $^6$, which
approximates the average spectral index of x-ray for FRSQs. While for BL
Lacs, the average spectral index at millimeter-submillimeter wavelength is
small $^6$.
The
approximate spectral index and the strong
correlation between the emission at the millimeter wavelength and the emission
of x-ray support
the view that the x-ray is mainly produced via SSC process
for EGRET QSOs. Seed photons that are scattered to x-ray regime probably
originate in
the millimeter region. For a large sample, Bloom \& Marscher $^1$$^9$  did not
find a correlation between x-ray flux and VLBI core flux at 10.7 GHz. The
VLBI core is
optically thick and the dependence of the VLBI core flux on the Doppler
factor $\delta$ is different from that of the x-ray emission.
In contrast, the emission at millimeter
wavelengths is optically thin and has the same dependence on the
Doppler factor $\delta$ as the x-ray emission, so, it is not difficult
to understand a correlation between the emissions at millimeter wavelengths
and the x-ray emissions, which might be attributed to SSC. It should also be
noted that the emissions at millimeter wavelengths come from a more
compact region  at lower radio frequency band, so,
the correlation
between the $\gamma$-ray emission and millimeter wavelength emission is
consistent with the result of Zhou et al $^2$$^0$, who found a strong correlation
between the $\gamma$-ray emission and the VLBI core flux at radio frequency.
Since the millimeter wavelength emissions are scattered to
x-ray regime,
the strong correlation between the $\gamma$-ray emission and the emission at
millimeter wavelength suggests
that the same electrons are responsible for both the $\gamma$-ray
emission and the millimeter wavelength emission. We should also expect a
correlation between the $\gamma$-ray
emission and the x-ray emission. Unfortunately, there are very little
corresponding x-ray observations during the Compton gamma-ray observatory
detection for most of the
sources in our sample.
A strong
correlation might appear when the data in both waveband are obtained almost
simultaneously. This is exactly the case of 3C279 $^1$$^5$, where
a dramatic x-ray flare is observed to occur almost simultanously with $\gamma$-ray
flare, suggesting a correlation.
The best way to establish the connection between the emission in
high energy band ($\gamma$-ray and x-ray) and that in lower frequencies is
through a series of simultaneous multi-waveband monitoring with high temporal
resolution.

We thank the  support from the  National Natural Science
Foundation of China under Grant No. 19703002 and Pandeng Plan.

\begin{figure}
 \centerline{\psfig{figure=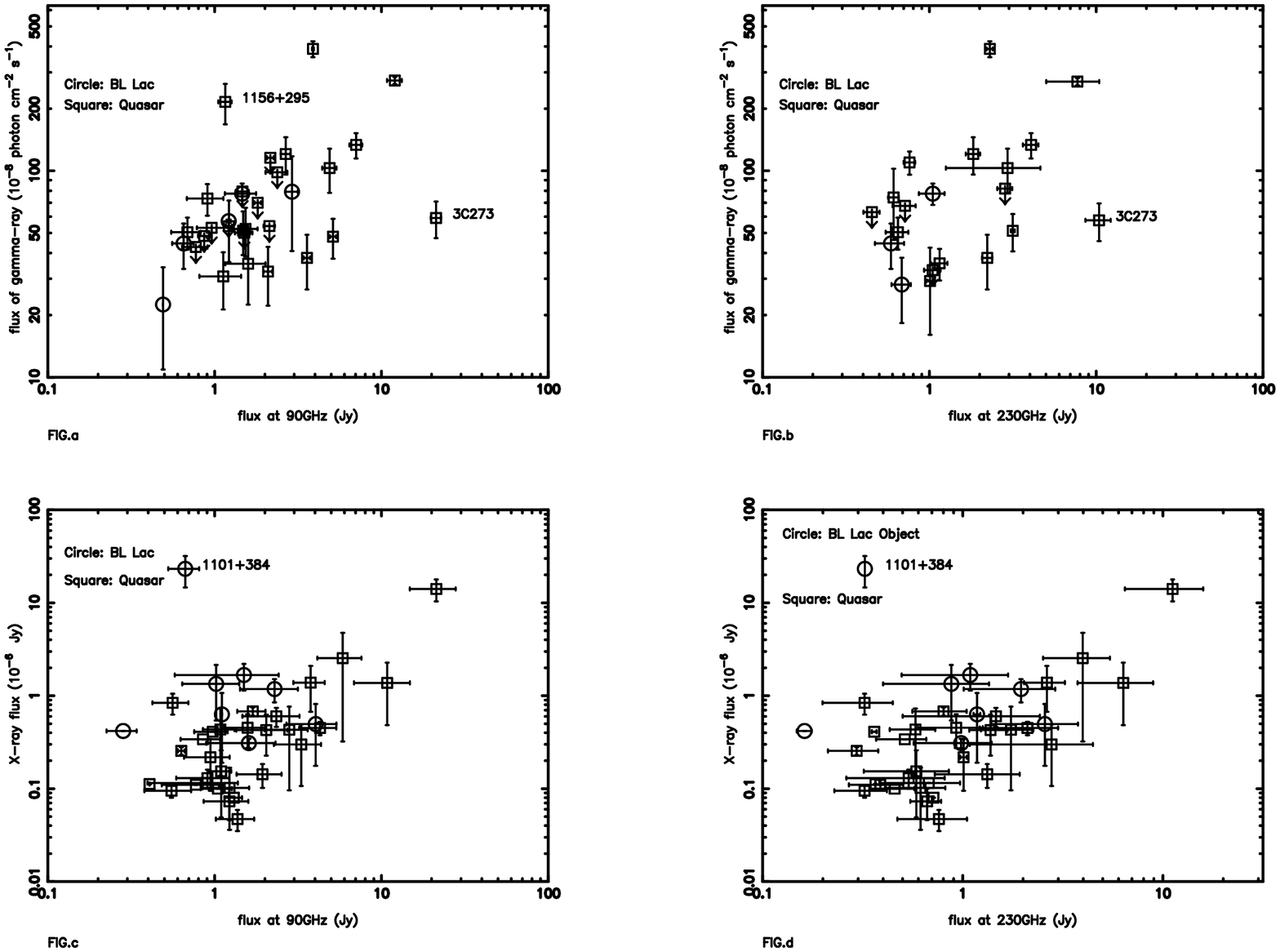,width=15.0cm,height=15.0cm,angle=-90}}
\caption[]
{The relation between millimeter wavelenth emissions and
high-energy emissions. Arrows in Figs. 1(a) and 1(b) denote the upper
limits of fluxes.} 
\end{figure}

\newpage
{\bf Reference.}
\vskip 1mm
$^1$ R. Mukherjee et al., Astrophys. J. 490 (1997) 116\\
$^2$ E. Valtaoja and H. Ter$\ddot{a}$sranta, Astron.Astrophys. 120 (1996) 491\\
$^3$ M. Tornikoski et al., Astron.Astrophys Suppl. 116 (1996) 157\\
$^4$ H. Steppe, et al., Astron. Astrophys, Suppl. 113(1995) 409\\
$^5$ H.P. Reuter et al., Astron. Astrophys, Suppl. Ser. 122 (1997) 271\\
$^6$ S.D. Bloom et al., Astron. J. 108 (1994) 398\\
$^7$ W.K. Gear et al, Mon.Not.R.Astron.Soc. 267 (1994) 167\\
$^8$ L. Dondi and G. Ghisellini, Mon.Not.R.Astron.Soc. 273 (1995) 583\\
$^9$ A. Comastri, G. Fossati, G. Ghisellini and S. Molendi, Astrophys. J. 480 (1997) 534\\
$^1$$^0$ G. Ghisellini, P. Padovani, A. Celotti and L. Maraschi, Astrophys. J. 407 (1993) 65\\
$^1$$^1$ B.J. Wilkes et al., Astrophys. J. Suppl. 92 (1994) 53\\
$^1$$^2$ G. Lamer, H. Brunner and  R. Staubert, Astron. Astrophys.
311 (1996) 384\\
$^1$$^3$ Y.F. Zhang et al., Astrophys.J. 432 (1994) 91\\
$^1$$^4$ D.M. Worrall and B.J. Wilkes, Astrophys.J. 360 (1990) 396\\
$^1$$^5$ A.E. Wehrle et al., Astrophys. J. 497 (1998) 178\\
$^1$$^6$ M. Pohl et al., Astron. Astrophys. 303 (1995) 383\\
$^1$$^7$ C.G. Bower at al., Astrophys. J. 484 (1997) 118\\
$^1$$^8$ I.M. McHardy et al., in Multi-Wavelength Continuum Emission of AGN,
edited by T.J.-L. Courvoisier and  A. Blecha. (Kluwer, Dordrecht), 1994, P.155\\
$^1$$^9$ S.D. Bloom and A.P. Marscher, Astrophys. J. 366 (1991) 16\\
$^2$$^0$ Y.Y. Zhou, Y.J. Lu, T.G. Wang, K.N. Yu, and E.C.M. Young,
Astrophys. J. 484 (1997) L47\\

\vskip 2.0cm
\end{document}